%% file: main.tex
\documentclass[onecolumn,pdflatex,final]{IEEEtran}
\usepackage{caption}
\usepackage{subcaption}  
\usepackage{algorithm}
\usepackage{algpseudocode}
\usepackage{algorithm} 
\usepackage[table]{xcolor}
\usepackage{longtable}
\usepackage{graphicx}
\usepackage{titlesec}
\usepackage{xcolor}
    \titleformat{\section}
    {\color{black}\normalfont\Large\bfseries}
    {\thesection}{1em}{}
    \titleformat{\subsection}
    {\color{black!70!black}\normalfont\large\bfseries}
    {\thesubsection}{1em}{}
    \usepackage[inline]{showlabels}
    
    \usepackage[colorlinks=true,linkcolor=black,citecolor=blue,filecolor=magenta,urlcolor=cyan,pdftex]{hyperref}
    \RequirePackage{xcolor}
    \hypersetup{
    colorlinks=true,
    linkcolor=blue,
    urlcolor=orange,
    }
    \usepackage{bookmark}

\usepackage[T1]{fontenc}
\usepackage{lmodern} 
\usepackage{amssymb,stmaryrd,amsmath,amsfonts,rotating}
\usepackage{amsmath,amsthm,amssymb,cases}
\usepackage{tikz}
\usetikzlibrary{arrows}
\usetikzlibrary{tikzmark}
\usetikzlibrary{matrix}
\usetikzlibrary{positioning}
\usepackage{algpseudocode}
\usepackage{amsmath,cases}
\usepackage{bbm}
\usepackage{amsthm}
\usepackage{booktabs}
\usepackage{xcolor}
\usepackage{amssymb}
\usepackage{bm}
\usepackage{thmtools}

\theoremstyle{definition}
\declaretheoremstyle[style=definition,qed=\openbox,]{ppstyle}

\usepackage{mathtools}
\mathtoolsset{showonlyrefs}

\input{notation}

\ifCLASSINFOpdf
\else
\fi
\begin{document}
%
\title{Sharp Error-Rate Transitions in Quantum QC-LDPC Codes under Joint BP Decoding}
\author{
\IEEEauthorblockN{Daiki Komoto and Kenta Kasai}
\IEEEauthorblockA{\\Institute of Science Tokyo\\
Email: $\mathtt{\{komoto.d.cdb8@m,kenta@ict.eng\}.isct.ac.jp}$}
}
\maketitle

\begin{abstract}
In this study, we report that quantum quasi-cyclic low-density parity-check codes decoded via joint belief propagation (BP) exhibit steep error-rate curves, despite the presence of error floors.  
To the best of our knowledge, this is the first observation of such threshold-like behavior for quantum LDPC codes with non-vanishing coding rate, excluding those decoded with non-binary BP decoders.  
Moreover, we find that dominant error events contributing to the error floor typically involve only a small number of bits.  
These findings suggest that the error floor is caused by trapping sets—specific subgraph structures in the Tanner graph—and indicate that identifying and avoiding such structures may lead to further reduction of the error floor.
\end{abstract}
\IEEEpeerreviewmaketitle

\section{Introduction}
A quantum error correction scheme exhibiting a steep frame error rate curve approaching the hashing bound and a low error floor has been reported in~\cite{komoto2024quantum}.
This decoding method utilizes a non-binary low-density parity-check (LDPC) decoder to decode quantum LDPC codes with computational complexity proportional to the code length.
However, the decoding complexity scales quadratically with the size of the finite field.

Classical coding theory is well established, and efficient coding schemes have been developed that achieve theoretical limits. 
These schemes exhibit steep reductions in error rate and low error floors as the code length increases.

To the best of the authors' knowledge, apart from the code proposed in~\cite{komoto2024quantum}, no quantum LDPC codes with non-vanishing coding rates have been reported to exhibit steep scaling behavior. 
In contrast, threshold phenomena have been extensively studied for quantum codes with vanishing rates~\cite{deMartiiOlius2024decodingalgorithms}, 
underscoring a fundamental distinction in decoding behavior between finite-rate and vanishing-rate quantum codes. 
This disparity suggests the need for further analysis to elucidate the mechanisms behind sharp transitions in the finite-rate regime.
One related exception is~\cite{gokduman2025isit}, which demonstrates that a certain class of LDPC codes achieves steep performance curves over the quantum erasure channel. 
While the channel model may not directly reflect realistic noise, the result provides a valuable theoretical benchmark.
Another notable example is~\cite{4957637}, which reports that a rate-1/9 quantum turbo code exhibits scalable waterfall behavior with sharp error-rate transitions, further supporting the feasibility of such phenomena beyond LDPC-based constructions.

In this work, we investigate the decoding error rate of quasi-cyclic (QC-) LDPC codes decoded using joint belief propagation (BP).  
Rather, the primary contribution of this study is to show that even quantum QC-LDPC codes decoded with a binary decoder can exhibit a sharp transition in the \black{frame error rate (FER) and} bit error rate (BER).  
This observation suggests that such steep behavior is not exclusive to codes with high decoding complexity, but can also arise under relatively simple decoding strategies.

Our simulation results reveal that the \black{FER and} BER exhibit a steep transition as the physical error rate decreases.  
\black{In contrast, such behavior was not observed for codes with small row weights.}  
This discrepancy suggests that most residual errors contributing to the error floor are composed of only a small number of bit errors.  
These findings indicate that certain subgraph structures—commonly referred to as \emph{trapping sets}—in the Tanner graph may be responsible for the error floor behavior.  
By identifying and understanding such structures, we aim to pave the way for future code designs that can avoid them and thus improve low-error performance.

\section{Code Construction and Decoding Method}

In this study, we employ quantum QC-LDPC codes.  
These codes are constructed from a pair of orthogonal parity-check matrices $(H_X,H_Z)$, each composed of $J \times L$ circulant permutation matrices (CPMs) of size $P$.  
An example of the QC-LDPC code used in this paper with $J=3$ and $L=8$ is given below.  
The exponent matrices \( E_X \) and \( E_Z \), corresponding to \( H_X \) and \( H_Z \), are given as follows:
\begin{align}
    E_X &= \left[
    \begin{array}{cccc|cccc}
        \textcolor{white}{-}2^
        0 & \textcolor{white}{-}2^1 & \textcolor{white}{-}2^2 & \textcolor{white}{-}2^3 & \textcolor{white}{-}2^4 & \textcolor{white}{-}2^5 & \textcolor{white}{-}2^6 & \textcolor{white}{-}2^7 \\
        \textcolor{white}{-}2^3 & \textcolor{white}{-}2^0 & \textcolor{white}{-}2^1 & \textcolor{white}{-}2^2 & \textcolor{white}{-}2^7 & \textcolor{white}{-}2^4 & \textcolor{white}{-}2^5 & \textcolor{white}{-}2^6 \\
        \textcolor{white}{-}2^2 & \textcolor{white}{-}2^3 & \textcolor{white}{-}2^0 & \textcolor{white}{-}2^1 & \textcolor{white}{-}2^6 & \textcolor{white}{-}2^7 & \textcolor{white}{-}2^4 & \textcolor{white}{-}2^5
    \end{array}
    \right] \\
    E_Z &= \left[
    \begin{array}{cccc|cccc}
        -2^4 & -2^7 & -2^6 & -2^5 & -2^0 & -2^3 & -2^2 & -2^1 \\
        -2^5 & -2^4 & -2^7 & -2^6 & -2^1 & -2^0 & -2^3 & -2^2 \\
        -2^6 & -2^5 & -2^4 & -2^7 & -2^2 & -2^1 & -2^0 & -2^3 
    \end{array}
    \right]
\end{align}
This code is a natural extension of the construction in~\cite{komoto2025explicit} to column weight three.
The construction can be similarly defined for other values of $L$.
By appropriately choosing the sub-blocks, the girth can be made equal to~6.

\begin{figure*}[t]
  \centering
  \begin{minipage}{0.49\linewidth}
    \centering
    \includegraphics[width=\linewidth]{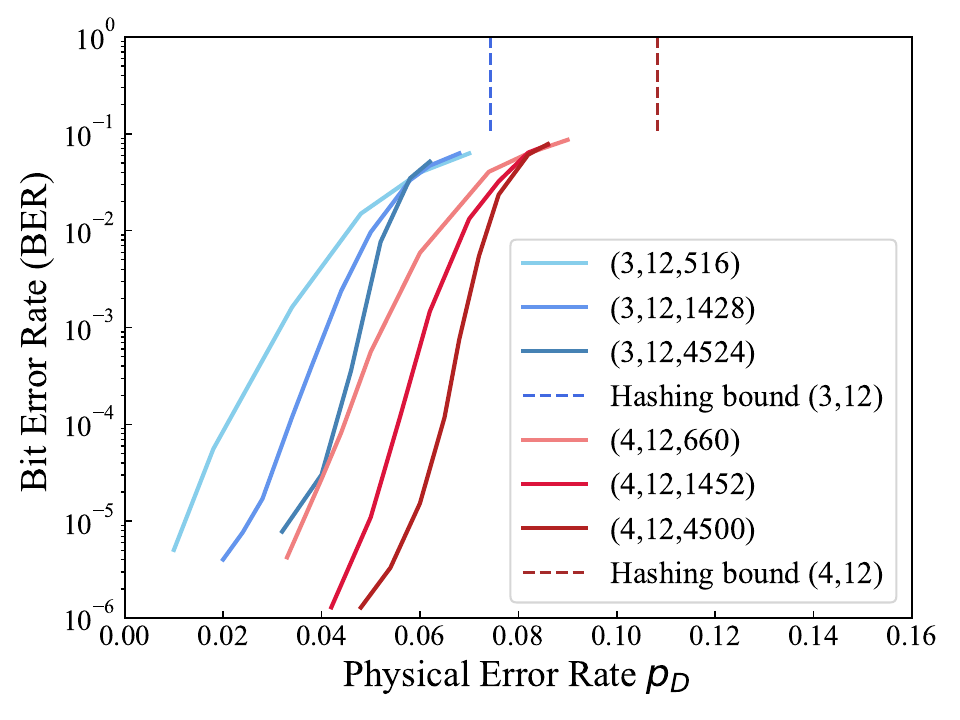}
  \end{minipage}
  \begin{minipage}{0.49\linewidth}
    \centering
    \includegraphics[width=\linewidth]{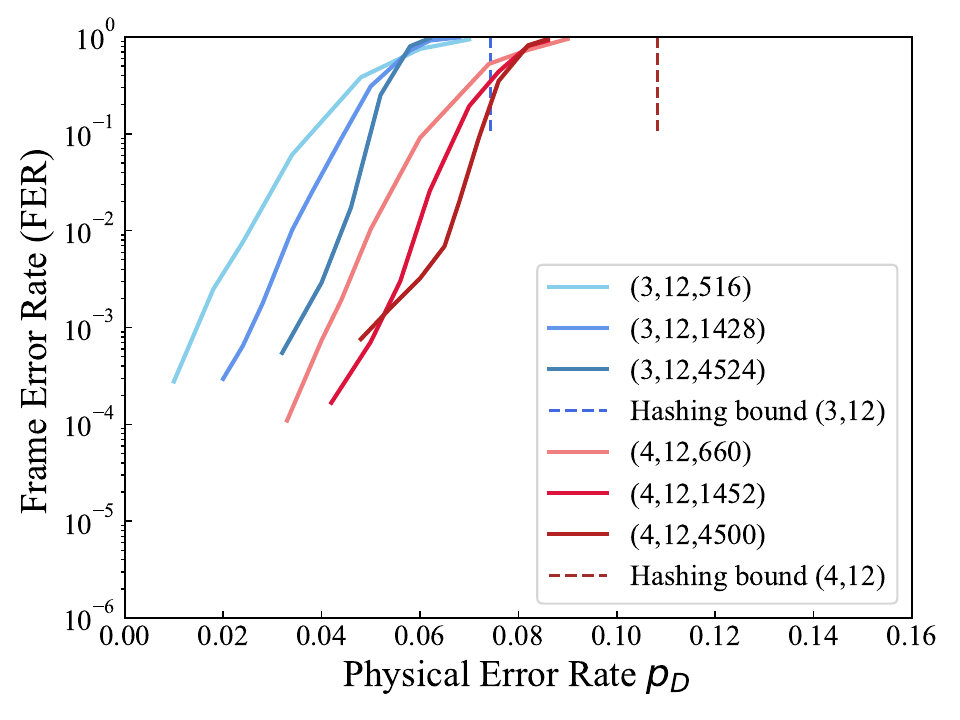}
  \end{minipage}
  \caption{\black{BER and  FER vs. physical error rate \( p_D \) for quantum QC-LDPC codes with \( (J, L, n) \).}}
  \label{fig:pDvsBER}
\end{figure*}

Accordingly, both $H_X$ and $H_Z$, which are orthogonal to each other, have column weight $J$ and row weight $L$.  
The code length is given by $n = PL$.
 The quantum coding rate \( R \) is defined as \( R = 1 - (\text{rank}(H_X) + \text{rank}(H_Z)) / n \). 

The quantum LDPC codes are decoded using a joint BP decoding algorithm.  
Let \( \underline{x} \) and \( \underline{z} \) denote the true noise, and \( \hat{\underline{x}} \) and \( \hat{\underline{z}} \) denote the estimated noise.  
Given the input syndrome \( (\underline{s},\underline{t}) \), the decoder simultaneously estimates \( \hat{\underline{x}} \) and \( \hat{\underline{z}} \) so as to approximate  maximization of the posterior probability \( p(\underline{x},\underline{z} \mid \underline{s},\underline{t}) \).  


The decoding algorithm terminates and outputs the current estimate \( (\hat{\underline{x}},\hat{\underline{z}}) \) when either of the following conditions is met:  
 \( (H_Z \hat{\underline{x}}, H_X \hat{\underline{z}}) = (\underline{s},\underline{t}) \) or the number of iterations reaches a preset maximum.
Decoding is defined to be successful if and only if  
\( \hat{\underline{x}} + \underline{x} \in C_X^\perp \) and  
\( \hat{\underline{z}} + \underline{z} \in C_Z^\perp \).  
Here, the term ``error floor'' refers to the flattening of the performance curve at low physical error rates.  

The FER is the fraction of trials that fail the stabilizer-equivalence success condition above. For the BER statistic, a successful trial contributes zero, including when its residual is a nonzero stabilizer. On a failed trial, the number of decoded bit errors is the number of indices \(i=0,\ldots,n-1\) such that \(x_i\ne\hat{x}_i\) or \(z_i\ne\hat{z}_i\).  
This BER statistic does not have a direct physical interpretation; rather, it quantifies the residual support on failed trials from a coding-theoretic perspective. The BER is the average of this count over all trials, divided by the code length \(n\). Thus FER uses equivalence modulo stabilizers and is not defined by exact equality of the sampled and estimated binary error vectors.

In classical LDPC codes, a sharp threshold phenomenon is observed in \black{both FER and BER}.  
In contrast, to the best of the authors' knowledge, such threshold behavior has not been reported for quantum LDPC codes, except for the code proposed in~\cite{komoto2024quantum}.  
The main contribution of this paper is to report that even quantum QC-LDPC codes decoded using decoder for binary codes can exhibit a sharp transition.

\section{Results and Discussion}


We assume a depolarizing channel with physical error rate \( p_D \).  
We evaluated the \black{FER and} BER performance of the joint BP decoder for the quantum QC-LDPC codes defined in the previous section, under several $(J, L, n)$ parameter settings.  
Figure~\ref{fig:pDvsBER} shows the quantum error correction performance obtained through simulations.
The figure also includes the hashing bound for the code rate \( R = 1 - 2J/L \) as a reference for performance evaluation.

As the code length increases, the \black{FER and BER} curves become steeper.  
\black{
To the best of the authors' knowledge, except for the code proposed in~\cite{komoto2024quantum}, no quantum error-correcting codes with non-vanishing coding rates have been reported to exhibit sharp transitions in FER.
}
We also observe that a lower coding rate leads to a larger gap from the hashing bound.  
This phenomenon is also known in classical binary LDPC codes.  
Furthermore, the error floor becomes more prominent as the code length increases. 


\black{
For the quantum LDPC codes used in our simulations, the dominant error patterns contributing to the deepest error floor—i.e., \( \hat{\underline{x}} + \underline{x} \) and \( \hat{\underline{z}} + \underline{z} \)—consisted mostly of a small number of bits.  
More precisely, 
for the \( (4,12,1452) \) code, 97\% of the observed errors in the error floor region involved no more than \( 3L \) bits.  
In addition, 
for the \( (4,12,4500) \) code, 98\% of the observed errors were confined to \( 2L \) bits or fewer.  
This suggests that the decoder is trapped by specific small structures in the Tanner graph of the quantum LDPC code.
}

\section{Conclusion and Future Work}

In this study, we demonstrated that quantum QC-LDPC codes decoded using joint belief propagation exhibit steep FER and BER curves.  

In our recent work~\cite{kasai2025degeneracy}, the structure of trapping sets responsible for the error floor was clarified for permutation-matrix-based quantum LDPC codes with column weight \( J = 2 \).  
Building on this analysis, we also proposed a post-processing technique applied after joint BP decoding, which was shown to significantly reduce the error floor.

It is expected that similar post-processing approaches can be extended to the case of \( J \ge 3 \) and large row weight \( L \), by applying analogous techniques to those developed in~\cite{kasai2025degeneracy}.

Based on these findings, two main directions remain for future work.  
First, further reduction of the error floor for large values of \( J \) and \( L \) through the development of suitable post-processing algorithms.  
Second, the realization of quantum error correction with low error floors and performance approaching the hashing bound by leveraging spatially coupled code constructions.

\bibliographystyle{IEEEtran}
\bibliography{main}
\end{document}

%% file: notation.tex
\usepackage{bbm}
\usepackage{mathrsfs}

\makeatletter
\newcommand\RedeclareMathOperator{%
  \@ifstar{\def\rmo@s{m}\rmo@redeclare}{\def\rmo@s{o}\rmo@redeclare}%
}
\newcommand\rmo@redeclare[2]{%
  \begingroup \escapechar\m@ne\xdef\@gtempa{{\string#1}}\endgroup
  \expandafter\@ifundefined\@gtempa
     {\@latex@error{\noexpand#1undefined}\@ehc}%
     \relax
  \expandafter\rmo@declmathop\rmo@s{#1}{#2}}
\newcommand\rmo@declmathop[3]{%
  \DeclareRobustCommand{#2}{\qopname\newmcodes@#1{#3}}%
}
\@onlypreamble\RedeclareMathOperator
\makeatother

\definecolor{grey}{rgb}{0.7, 0.75, 0.71}

\def\dark-red#1{\textcolor[rgb]{0.7,0.0,0.0}{#1}}
\def\black#1{\textcolor[rgb]{0.0,0.0,0.0}{#1}}
\definecolor{amber}{rgb}{1.0, 0.75, 0.0}

\def\...{\dotsc}

\def\intT2{\int_{-T/2}^{T/2}}

\def\sumi1n{\sum_{i=1}^{n}}
\def\sumi1N{\sum_{i=1}^{N}}
\def\sumi0N--{\sum_{i=0}^{N-1}}

\def\ccc{\cdots}

\def\=def{\overset{\text{\small def}}{=}}

\DeclarePairedDelimiterX{\inp}[2]{\langle}{\rangle}{#1, #2}

\def\<{\langle}
\def\>{\rangle}

 \def\mat4#1#2#3#4{
\begin{pmatrix}
 #1&\ccc&#2\\
 \vdots&&\vdots\\
 #3&\ccc&#4
\end{pmatrix}}

\def\0sf{\mathsf{0}}
\def\1sf{\mathsf{1}}

\def\0BS{\boldsymbol{0}}
\def\1BS{\boldsymbol{1}}

\def\0B{\mathbf{0}}
\def\1B{\mathbf{1}}

\def\0H{\hat{0}}
\def\1H{\hat{1}}

\def\+TT{\texttt{+}}
\def\-{\texttt{-}}
\def\+KB{|+\> \<+|}
\def\-KB{|-\> \<-|}

\def\q0{|0\>}

\def\0U{\underline{0}}
\def\1U{\underline{1}}

\def\0UH{\underline{\0H}}
\def\1UH{\underline{\1H}}

\RedeclareMathOperator{\Im}{Im}

\def\thepage{\arabic{page}}